\documentclass[conference,a4paper]{APSIPA2021}
\usepackage{amsmath}
\usepackage{graphicx}
\usepackage{multirow}
\usepackage{threeparttable}
\usepackage[backend=biber,style=ieee,]{biblatex}
\usepackage{geometry}
\usepackage{fancyhdr}

\fancypagestyle{firststyle}{
  \fancyhf{}
  \fancyhead[C]{2023 Asia Pacific Signal and Information Processing Association Annual Summit and Conference (APSIPA ASC)}
}

\usepackage{bm}
\DeclareMathOperator*{\argmax}{arg\,max}
\DeclareMathOperator*{\argmin}{arg\,min}
\newcommand{\vecw}{\bm{w}}
\newcommand{\vecs}{\bm{s}}
\newcommand{\vecn}{\bm{n}}
\newcommand{\vecx}{\bm{x}}
\newcommand{\phix}{\bm{\Phi}_{x}}
\newcommand{\phis}{\hat{\bm{\Phi}}_{s}}
\newcommand{\phin}{\hat{\bm{\Phi}}_{n}}
\newcommand{\maskset}{\mathcal M}

\newcommand{\htp}[1]{{#1}^{\rm H}}
\newcommand{\tp}[1]{{#1}^{\rm T}}
\newcommand{\conj}[1]{\overline{#1}}
\newcommand{\average}[1]{\left\langle#1\right\rangle}
\newcommand{\abs}[1]{\left|#1\right|}
\newcommand{\reffig}[1]{Fig.~\ref{#1}}
\newcommand{\reftable}[1]{Table~\ref{#1}}
\newcommand{\refeqn}[1]{(\ref{#1})}
\newcommand{\refsec}[1]{Section~\ref{#1}}
\newcommand{\nop}[1]{}

\begin{document}

\title{
Is the Ideal Ratio Mask Really the Best? --- Exploring the Best Extraction Performance and Optimal Mask of Mask-based Beamformers
}

\author{
\authorblockN{
Atsuo Hiroe\authorrefmark{1},
Katsutoshi Itoyama\authorrefmark{1}\authorrefmark{2}, and
Kazuhiro Nakadai\authorrefmark{1}
}

\authorblockA{
\authorrefmark{1}
Department of Systems and Control Engineering, School of Engineering, Tokyo Institute of Technology, Tokyo, Japan}

\authorblockA{
\authorrefmark{2}
Honda Research Institute Japan Co., Ltd., Saitama, Japan\\
E-mail: \{hiroe,itoyama,nakadai\}@ra.sc.e.titech.ac.jp}
}

\maketitle
\pagestyle{fancy}

\begin{abstract}
This study investigates mask-based beamformers (BFs), which estimate filters to extract target speech using time-frequency masks. Although several BF methods have been proposed, the following aspects are yet to be comprehensively investigated. 1) Which BF can provide the best extraction performance in terms of the closeness of the BF output to the target speech? 2) Is the optimal mask for the best performance common for all BFs? 3) Is the ideal ratio mask (IRM) identical to the optimal mask? Accordingly, we investigate these issues considering four mask-based BFs: the maximum signal-to-noise ratio BF, two variants of this, and the multichannel Wiener filter (MWF) BF. To obtain the optimal mask corresponding to the peak performance for each BF, we employ an approach that minimizes the mean square error between the BF output and target speech for each utterance. Via the experiments with the CHiME-3 dataset, we verify that the four BFs have the same peak performance as the upper bound provided by the ideal MWF BF, whereas the optimal mask depends on the adopted BF and differs from the IRM. These observations differ from the conventional idea that the optimal mask is common for all BFs and that peak performance differs for each BF. Hence, this study contributes to the design of mask-based BFs.
\end{abstract}

\section{Introduction}\label{sec:intro}

Target speech extraction is effective in improving speech intelligibility in telecommunication systems and the performance of automatic speech recognition systems~\cite{Chen2018-rl}. Therefore, beamformers (BFs) are utilized to avoid nonlinear distortions such as musical noises and spectral distortions~\cite{Mizumachi2016-jd, Mizumachi2019-hi}. In the last decade, combined frameworks comprising BFs and deep neural networks (DNNs), referred to as mask-based BFs, have been proposed \cite{Heymann2016-eb, Heymann2016-sy, Erdogan2016-jc}. In these frameworks, DNNs generate one or two time-frequency (TF) masks corresponding to the target, interferences, or both to inform the BF of the sound to be enhanced or suppressed. Afterward, the BF estimates a filter for extracting the target using masks. For filter estimation, the following BF methods are adopted:
1) maximum signal-to-noise ratio (max-SNR) or generalized eigenvalue (GEV) BF~\cite{Heymann2016-eb, Heymann2016-sy, Drude2019-so},
~2) minimum variance distortionless response (MVDR) BF~\cite{Heymann2016-sy, Erdogan2016-jc, Souden2010-kp}, and
3) multichannel Wiener filter (MWF) BF~\cite{Stenzel2013-hw, Nugraha2016-rn, Pfeifenberger2017-zp}.

Several mask types have been examined to achieve improved extraction performance.
Initially, ideal binary masks (IBMs) were employed to train DNNs for mask generation \cite{Heymann2016-eb, Heymann2016-sy}. Subsequently, ideal ratio masks (IRMs) have been utilized \cite{Erdogan2016-jc,Wang2018-yy,Pfeifenberger2017-zp}.

Overviewing these studies, we concluded that the following aspects are yet to be comprehensively investigated:
\begin{enumerate}
    \item Which BF can achieve the best performance if an optimal mask is provided for each BF method?
    \item Is an optimal mask common for all BF methods?
    \item Are conventional ideal masks such as IRMs identical to an optimal mask?
\end{enumerate}
Regarding these aspects, several studies such as \cite{Heymann2016-sy,Boeddeker2018-ww}, and \cite{Wang2018-tl} considered that the mask optimal for the single-channel TF masking \cite{Wang2014-gj,Wang2018-yy} should commonly be optimal for all BF methods. However, this assumption was not verified in these studies. Moreover, they compared multiple BF methods that employed the same mask in terms of extraction performance. However, these results do not answer the first aspect unless the optimal mask is common.

The remainder of this paper is organized as follows. Sections~\ref{sec:related-works} and \ref{sec:framework} explain the related work and BF methods presented in this study, respectively. \refsec{sec:optimal-mask} examines the method for obtaining the optimal mask for each BF method. \refsec{sec:experiments} verifies the aforementioned points experimentally while \refsec{sec:discussion} discusses the experimental results. Finally, \refsec{sec:conclusions} concludes the study.

\section{Related work} \label{sec:related-works}
First, we overview the history of mask-based BFs. The max-SNR, MVDR, and MWF BFs employ the statistics of the target, interferences, or both. The statistics are called target (or speech) and interference (or noise) covariance matrices. Given that the accuracy of both statistics affects the extraction performance, estimating them accurately is a fundamental issue \cite{Araki2007-yp, Warsitz2007-fn}.
In \cite{Heymann2016-eb, Heymann2016-sy}, both statistics were computed using two binary masks, each of which represents periods when only the target or interferences are present, and the masks were generated using a properly trained DNN. This idea was first adopted for the max-SNR and MVDR BFs \cite{Heymann2016-eb, Heymann2016-sy, Erdogan2016-jc}, then applied to the MWF BF \cite{Heymann2018-mf}.

Second, we mention the studies that compare mask-based BFs. At least three methods have been adopted as aforementioned, and several studies have compared two or three of them, as presented in \reftable{tab:comparison}. However, the best-performing method depends on experimental setups. Moreover, it is reported in \cite{Heymann2018-mf} that the difference between the max-SNR and MWF BFs is marginal and the performance depends on the number of microphones used.

\begin{table}[b]
    \centering
    \caption{Studies comparing multiple BF methods. Methods in bold type represent those that performed the best in each study. In \cite{Heymann2018-mf}, the performance depends on the number of microphones. (SNR: Signal-to-noise ratio, PESQ: Perceptual evaluation of speech quality, WER: Word error rate)}
    \label{tab:comparison}

    \begin{tabular}{l|l|l}
    \hline
& Metric & Methods compared \\
    \hline
Heymann+16  \cite{Heymann2016-sy}   & SNR & {\bf Max-SNR}, MVDR \\
Boeddeker+18\cite{Boeddeker2018-ww} & SNR,PESQ & {\bf Max-SNR}, MVDR \\
Wang+18     \cite{Wang2018-tl}      & WER & Max-SNR, MVDR, {\bf MWF}\\
Heymann+18  \cite{Heymann2018-mf}   & WER & Max-SNR, MWF \\
Shimada+18  \cite{Shimada2018-uy}   & WER & MVDR, {\bf MWF} \\
\hline
    \end{tabular}
\end{table}

Third, we mention the mask types considered to be optimal for the mask-based BFs.
Initially, the IBMs were considered to be optimal \cite{Heymann2016-eb, Heymann2016-sy}; then, the IRMs were considered optimal \cite{Erdogan2016-jc,Tu2017-ey,Wang2018-yy,Tu2019-qp,Chakrabarty2019-th}.
These ideas were based on the findings in the single-channel TF masking \cite{Wang2014-gj,Wang2018-yy}. However, these studies did not investigate whether these findings really apply to the mask-based BFs or the optimal mask is common for any BF method.

\section{Framework of mask-based BFs} \label{sec:framework}
In this study, we define the best extraction performance as the BF output closest to the target in the TF domain, considering that a significant goal of BFs is to extract (or estimate) the target. Moreover, similar to conventional TF masks, we have the constraint that mask values are nonnegative and real-valued.

\begin{figure}[t]
  \centering
  \includegraphics[width=0.6\linewidth]{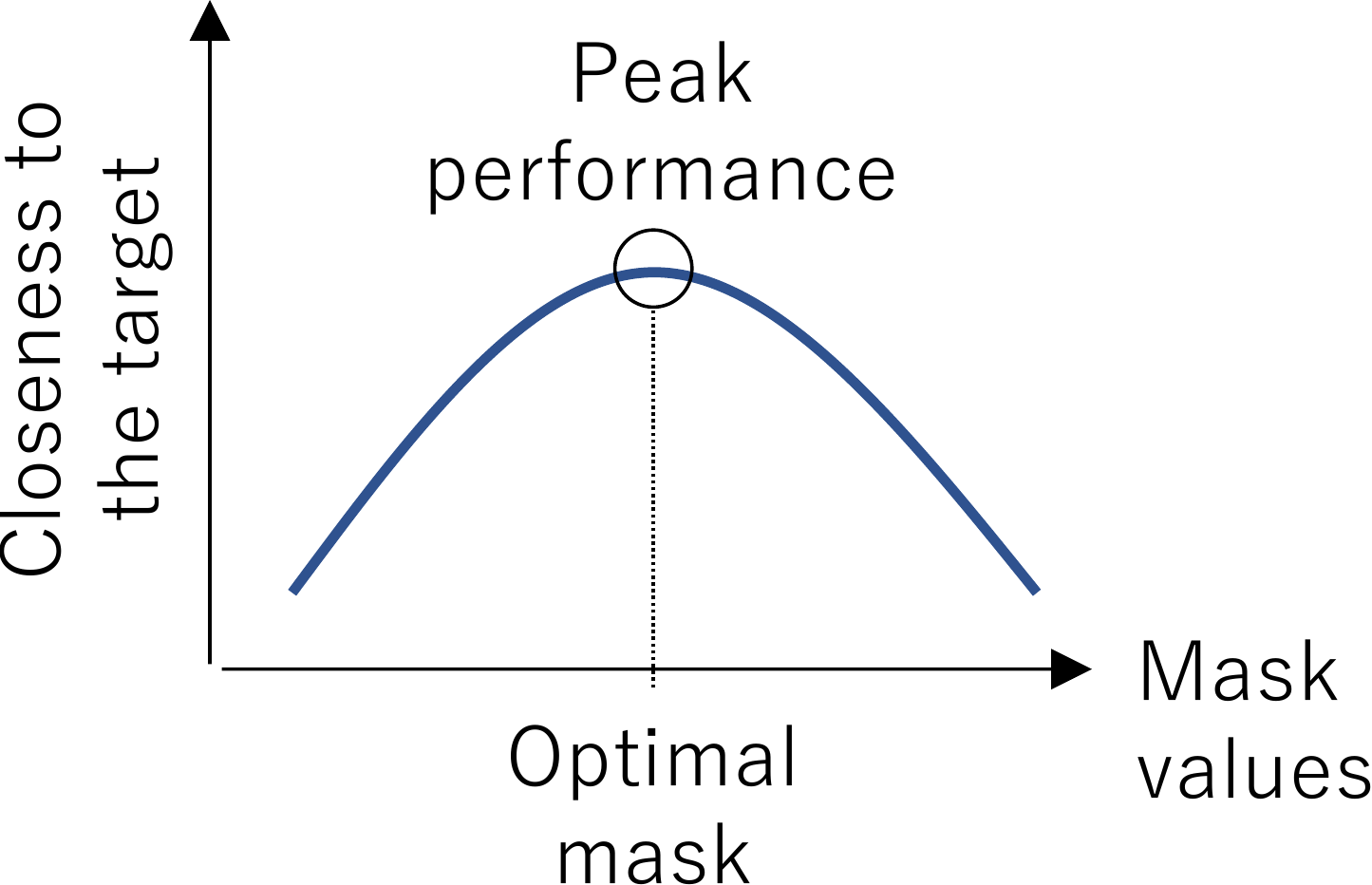}
  \caption{Conceptual plot of the relationship between the closeness of the BF output to the target and mask values}
  \label{fig:mask_curve}
\end{figure}

\reffig{fig:mask_curve} illustrates our idea. The vertical and horizontal axes indicate the closeness of the BF output to the target and the variation of the mask values, respectively. Although the mask values vary multidimensionally, this figure conceptually represents the variation as a single axis. In the mask-based BFs, the extraction performance should depend on the variation and exhibit the peak on a particular mask. We refer to this mask as the optimal one.
Considering that multiple BF methods and mask types are employed, we can rephrase the questions mentioned in \refsec{sec:intro} as follows:
  \begin{description}
    \item[\bf Issue 1:] \quad Which BF method has the highest peak, or are all peaks of the same height?
    \item[\bf Issue 2:] \quad Is the optimal mask common for all BF methods, or dependent on each one?
    \item[\bf Issue 3:] \quad Can the mask considered to be ideal achieve peak performance? More particularly, is the IRM optimal?
  \end{description}

\begin{figure}[t]
  \centering
  \includegraphics[width=0.6\linewidth]{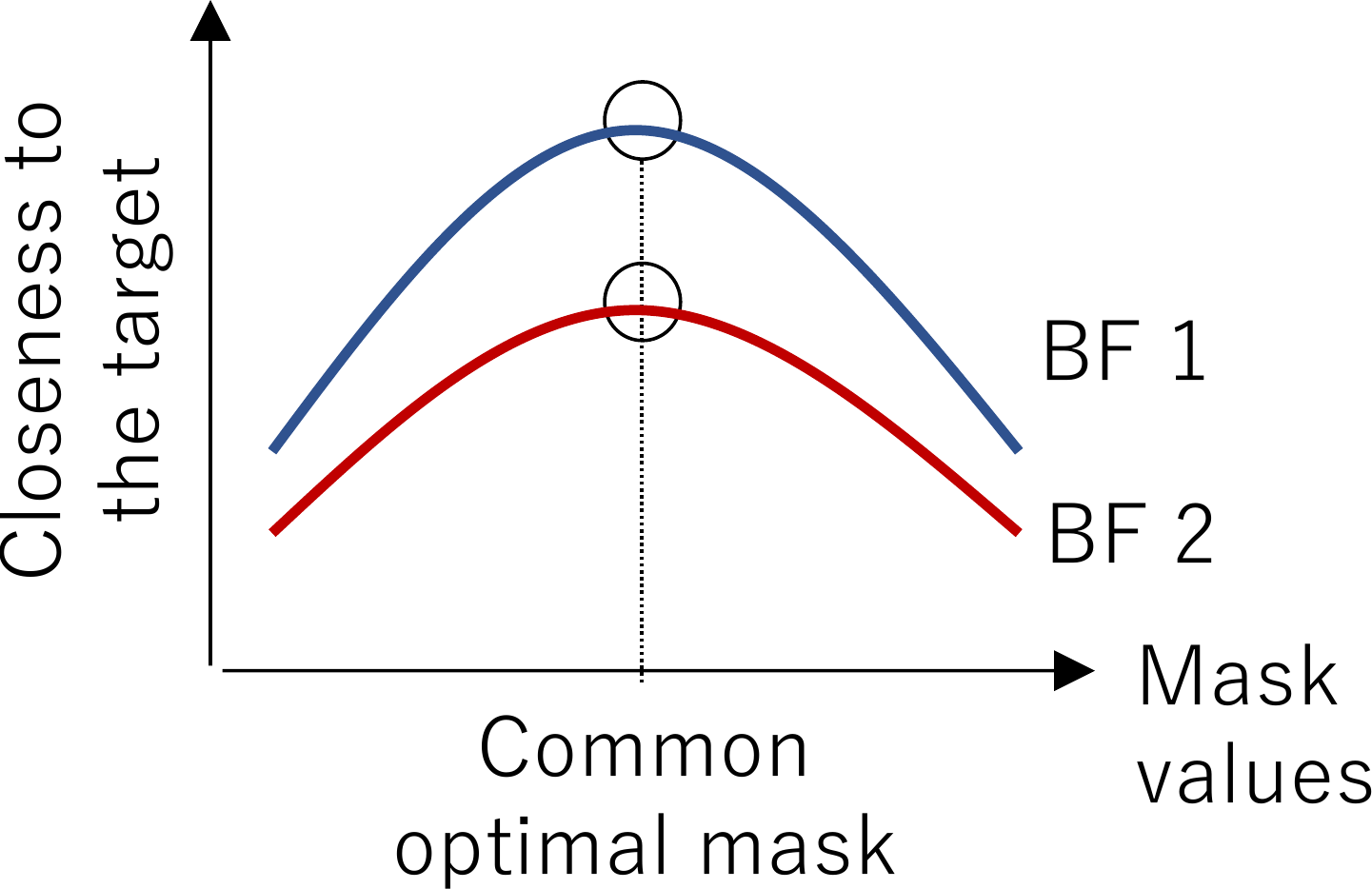}
  \caption{Conventional ideas on the peak performance and optimal mask for multiple BF methods}
  \label{fig:conventional_image}
\end{figure}

We assumed that the studies presented in \reftable{tab:comparison} considered these issues as in \reffig{fig:conventional_image}. This indicates that the mask mentioned in each study is optimal for multiple BF methods such as BFs 1 and 2, whereas each BF method demonstrates a different height of the peak. However, this idea is yet to be verified.

Considering that this study is the first to examine these issues, we focus on the following methods for simplicity:
\begin{itemize}
    \item Mask-based MWF BF, which employs a single mask
    \item Mask-based max-SNR BF, which utilizes two masks, and its variants that employ a single mask
\end{itemize}
We do not include the MVDR BF in this study, given that the method has another issue regarding the accuracy of the estimation of the steering vector \cite{Tu2018-ts}.

\subsection{Signal models}
In this study, all signals are in the TF domain. For simplicity, the frequency index is omitted, whereas the frame index $t$ is adopted. Let $\vecx(t)=\tp{[x_{1}(t),\ldots ,x_{N}(t)]}$ be an observation vector obtained with $N$ microphones. The observation $\vecx(t)$ can be expressed as the following summation:
\begin{equation}
    \vecx(t) = \vecs(t) + \vecn(t), \label{eqn:components-of-x}
\end{equation}
where $\vecs(t)=\tp{[s_1(1), \ldots, s_N(t)]}$ denotes the components arriving from the target and $\vecn(t)=\tp{[n_1(1), \ldots, n_N(t)]}$ represents the residuals called interferences. Using the observation $\vecx(t)$ and BF filter $\vecw$, the BF output $y(t)$ is expressed as
\begin{equation}
    y(t) = \htp{\vecw} \vecx(t). \label{eqn:bf-output}
\end{equation}

We use the following three covariance matrices:
\begin{eqnarray}
\phix &=& \average{\vecx(t)\htp{\vecx(t)}}_{t},
    \label{eqn:phix}\\
\phis &=& \average{m_{s}(t)\vecx(t)\htp{\vecx(t)}}_{t},
    \label{eqn:phis}\\
\phin &=& \average{m_{n}(t)\vecx(t)\htp{\vecx(t)}}_{t},
    \label{eqn:phin}
\end{eqnarray}
where $m_{s}(t)$ and $m_{n}(t)$ denote TF masks for the target and interferences, respectively, and $\average{\cdot}_{t}$ computes the average over $t$. Each mask comprises nonnegative real values. We refer to these matrices as observation, target, and interference covariance matrices, respectively. Unlike $\phix$, both $\phis$ and $\phin$ are estimated matrices computed from the masks and observations without using $\vecs(t)$ and $\vecn(t)$.

Moreover, consider ${\rm GEV_{max}}(\bm{A},\bm{B})$ and ${\rm GEV_{min}}(\bm{A},\bm{B})$ to be the eigenvectors corresponding to the maximum and minimum eigenvalues in the following GEV problem, respectively:
\begin{equation}
    \bm{Aw} = \lambda \bm{Bw}.
\end{equation}

\subsection{Mask-based and ideal MWF BFs}
\label{subsec:mwf}

The MWF BF is formulated as a problem of minimizing the mean square error (MSE) between the BF output and the corresponding reference $p(t)$ \cite{Wang2018-tl}:
\begin{eqnarray}
\vecw &=& \argmin_{\vecw} \average{\left|p(t) - \htp{\vecw}\vecx(t)\right|^{2}}_{t}
    \label{eqn:mwf-def} \\
        &=& \phix^{-1} \average{\vecx(t) \conj{p(t)}}_{t},
\end{eqnarray}
where $\conj{p(t)}$ denotes the conjugate of $p(t)$. The mask-based MWF BF employs the masked observation as the reference, and thus, corresponds to the case $p(t)=m_{s}(t)x_{k}(t)$ as
\begin{eqnarray}
\vecw_{\rm mwf} &=& \argmin_{\vecw} \average{\left|m_{s}(t)x_{k}(t) - \htp{\vecw}\vecx(t)\right|^{2}}_{t}
    \label{eqn:mask-mwf-def} \\
                &=& \phix^{-1} \average{m_{s}(t) \vecx(t) \conj{x_{k}(t)}}_{t},
    \label{eqn:mask-mwf}
\end{eqnarray}
where $x_{k}(t)$ denotes the observation obtained with the $k$th microphone.

Another significant variant is the ideal MWF BF, which provides the upper-bound extraction performance for all BFs \cite{Malek2020-nw}.
When $\vecs(t)$ in \refeqn{eqn:components-of-x} is known, the upper-bound extraction performance can be achieved using the MWF with $p(t)=\vecs_{k}(t)$.
The corresponding filter is obtained as
\begin{eqnarray}
\vecw_{\rm ideal} &=& \argmin_{\vecw} \average{\left|s_{k}(t) - \htp{\vecw}\vecx(t)\right|^{2}}_{t}
    \label{eqn:ideal-mwf-def} \\
                  &=& \phix^{-1} \average{\vecx(t) \conj{s_{k}(t)}}_{t}.
    \label{eqn:ideal-mwf}
\end{eqnarray}

Note that the ideal MWF BF is not a particular case of the mask-based one. This is because a mask value $m_{s}(t)$ constrained to be real-valued and nonnegative cannot render \refeqn{eqn:mask-mwf-def} equivalent to \refeqn{eqn:ideal-mwf-def}, whereas $m_{s}(t)$ that can take any complex values can. Therefore, it is not evident whether the mask-based MWF BF can achieve the same extraction performance as the ideal alternative.

\subsection{Max-SNR BF and its variants}
\label{subsec:max-snr}

The mask-based max-SNR BF is formulated as the following maximization problem \cite{Heymann2016-eb,Heymann2016-sy}:
\begin{eqnarray}
\vecw_{\rm snr}
    &=& \argmax_{\vecw} \frac{\htp{\vecw}\phis\vecw}{\htp{\vecw}\phin\vecw}
            \label{eqn:max-snr-ratio} \\
    &=& {\rm GEV_{max}}\left(\phis, \phin\right).
            \label{eqn:max-snr-gev}
\end{eqnarray}
Although this method originally utilizes two masks, we can derive two different variants that adopt a single mask by assuming the following relationship:
\begin{equation}
     \phis + \phin = \phix .
        \label{eqn:cov_summation}
\end{equation}
One variant utilizes a mask for the interferences  \cite{Warsitz2007-fn, Hiroe2022-rd}:
\begin{eqnarray}
\vecw_{\rm nor}
    &=& \argmin_{\vecw} \frac{\htp{\vecw}\phin\vecw}{\htp{\vecw}\phix\vecw}
            \label{eqn:min-nor-ratio} \\
    &=& {\rm GEV_{min}}\left(\phin, \phix\right).
        \label{eqn:min-nor-gev}
\end{eqnarray}
We refer to this as the minimum noise-to-observation ratio (min-NOR) BF. Similarly, we can derive the other, which employs a mask for the target and is expressed as
\begin{eqnarray}
\vecw_{\rm sor}
    &=& \argmax_{\vecw} \frac{\htp{\vecw}\phis\vecw}{\htp{\vecw}\phix\vecw}
            \label{eqn:max-sor-ratio} \\
    &=& {\rm GEV_{max}}\left(\phis, \phix\right).
        \label{eqn:max-sor-gev}
\end{eqnarray}
We refer to this as the maximum signal-to-observation ratio (max-SOR) BF.

Furthermore, even without assuming \refeqn{eqn:cov_summation}, the min-NOR and max-SOR BFs can be equivalent to each other in terms of the filter estimation if one mask is computed from the other using \refeqn{eqn:sum-is-constant}. 
\begin{equation}
    m_{s}(t) + m_{n}(t) = \alpha,
        \label{eqn:sum-is-constant}\\
\end{equation}
where $\alpha$ is a nonnegative value, such that all the mask values are nonnegative. For example, assigninig $m_{s}(t) = \alpha - m_{n}(t)$ to \refeqn{eqn:max-sor-ratio} and \refeqn{eqn:phis} results in the problem represented by \refeqn{eqn:min-nor-ratio} if $m_{n}(t)$ is nonnegative for all $t$.

For the max-SNR BF and its variants, the scales of the filter and BF outputs are uncertain, unlike the MWF BF. Hence, a post-process for determining the proper scales is required \cite{Araki2007-yp, Warsitz2007-fn}.
Considering that the formulation of these BFs differs from that of the MWF BF, it is not evident whether these BFs can achieve the same extraction performance as the ideal MWF BF.

\section{Obtaining the optimal mask} \label{sec:optimal-mask}
To explore the peak performance for each BF method, this study employs a bottom-up approach that obtains the optimal mask for each mixture of the target and interferences, instead of a priori deciding whether a particular mask type is optimal.

Let $\maskset$ be a set of mask values adopted in a BF. This set comprises $m_{s}(t)$, $m_{n}(t)$, or both for all $t$, depending on the BF method employed. We can formulate $\maskset$ as the solution to the problem of minimizing the following MSE:
\begin{equation}
\maskset = \argmin_{\maskset} \average{\left|s_{k}(t) - y(t)\right|^{2}}_{t},
\label{eqn:optimal-mask-def}
\end{equation}
where $s_{k}(t)$ and $y(t)$ denote the target included in the observation of the $k$th microphone and BF output, respectively.
To render the mask values nonnegative and to avoid both diverging the mask values and converging them to zero, we impose the following constraints on \refeqn{eqn:optimal-mask-def}:
\begin{eqnarray}
    m(t) &\geq& 0 \quad \mbox{\rm for all $t$}, \label{eqn:constraint-nonnegative} \\
    \average{m(t)^{2}}_{t} &=& 1, \label{eqn:constraint-unitvariance}
\end{eqnarray}
where $m(t)$ denotes $m_{s}(t)$ or $m_{n}(t)$.
In \refeqn{eqn:optimal-mask-def}, $y(t)$ is computed as follows. First, the BF filter $\vecw(t)$ is computed depending on the adopted BF method, then the $y(t)$ is computed using \refeqn{eqn:bf-output}. To determine the ideal scale of $y(t)$ independent of the BF method, we apply the following post-filtering process, referred to as the {\it ideal scaling}:
\begin{eqnarray}
    \gamma &=& \frac{\average{s_{k}(t)\overline{y(t)}}_{t}} {\average{|y(t)|^{2}}_{t}},
        \label{eqn:ideal-scaling}\\
    y(t) &\leftarrow& \gamma y(t) \label{eqn:apply-ideal-scale}.
\end{eqnarray}
Owing to the ideal scaling, the constraint represented as \refeqn{eqn:constraint-unitvariance} does not affect the scale of $y(t)$.

Because $\maskset$, $\vecw(t)$, and $y(t)$ depend on each other, $\maskset$ cannot be obtained explicitly. In contrast, we employ the iterative algorithm based on gradient descent.
Given that the mask estimation process adopts no explicit association between the masks and sources, the obtained masks such as $m_{s}(t)$ and $m_{n}(t)$ do not necessarily correspond to the sources such as the target and interferences.

Equations \refeqn{eqn:optimal-mask-def}, \refeqn{eqn:constraint-nonnegative}, and \refeqn{eqn:constraint-unitvariance} might seem to solve the same problem as the ideal MWF BF, which is represented as \refeqn{eqn:ideal-mwf-def}, based on the iterative algorithm and consequently achieve the same extraction performance as the ideal MWF BF. However, such a perspective is incorrect, as mentioned in Sections~\ref{subsec:mwf} and \ref{subsec:max-snr}.
Rather, the objective of this study is to examine to what extent the output of the mask-based BF can approach that of the ideal MWF BF if the mask best fits \refeqn{eqn:optimal-mask-def}.

\section{Experiments} \label{sec:experiments}

To clarify Issues 1--3 mentioned in \refsec{sec:framework}, we conducted the following experiments:
\begin{enumerate}
    \item Exploring the peak performance for each BF method
    \item Verifying whether the optimal mask is common for all the BF methods
    \item Examining whether the IRM can achieve the peak performance
\end{enumerate}
The following subsections first mention the dataset and system for these experiments, and then demonstrate the experimental results in order.

\subsection{Dataset and experimental system}
We employed the CHiME-3 simulated test set \cite{Barker2017-tr}, which comprises both 330 utterances from four speakers and four background (BG) noises. In this dataset, sound data were recorded at 16~kHz with six microphones attached to a tablet device. We generated the TF domain signals using the short-time Fourier transform with window and shift lengths of 1024 and 256, respectively.

\begin{figure}[t]
  \centering
  \includegraphics[width=0.9\linewidth]{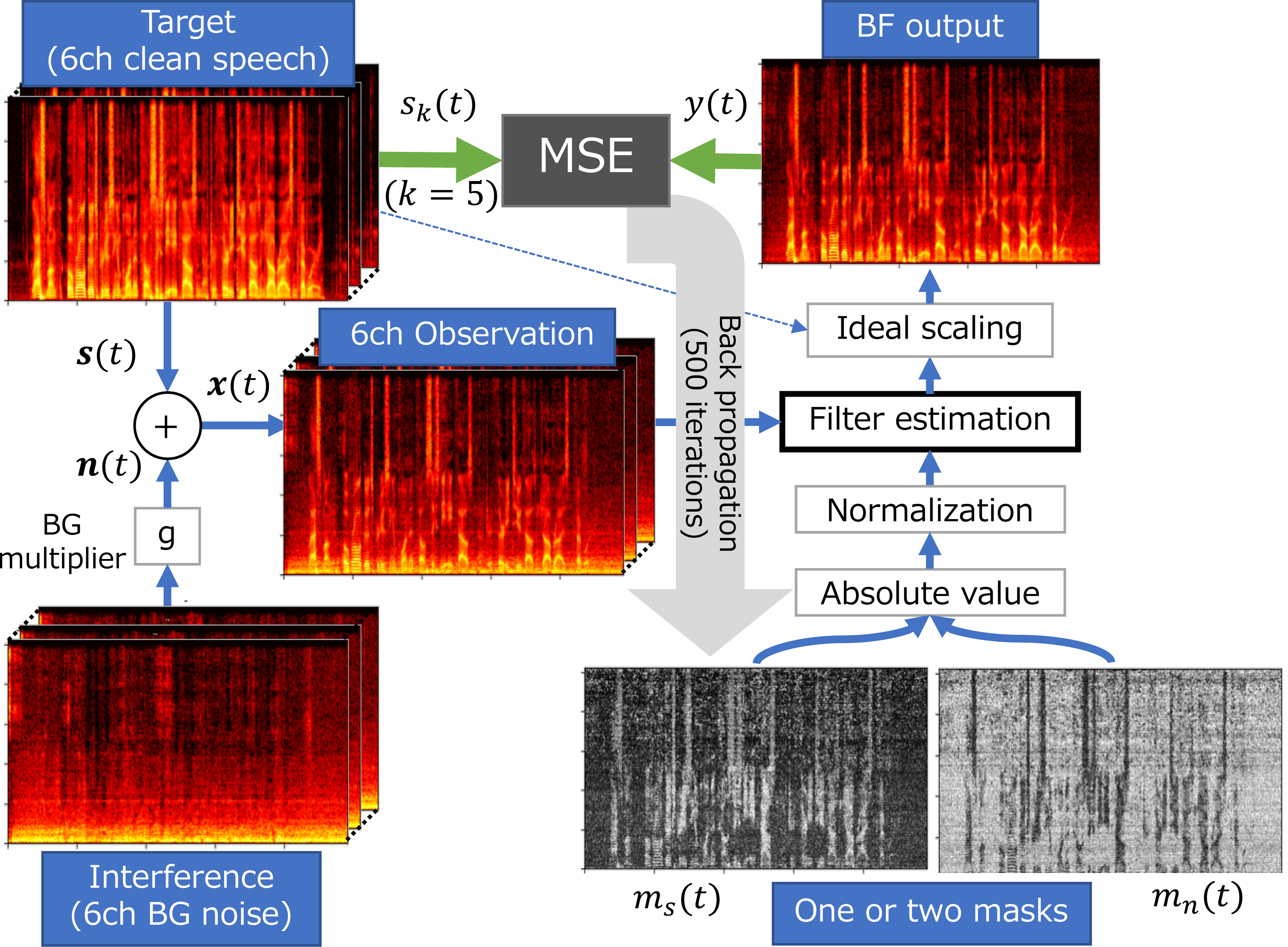}
  \caption{Configuration of the system obtaining or employing the optimal mask(s) for each utterance}
  \label{fig:system}
\end{figure}

\reffig{fig:system} illustrates the experimental system.
The modules labeled {\it Absolute value}, {\it Normalization}, and {\it Ideal scaling} correspond to \refeqn{eqn:constraint-nonnegative}, \refeqn{eqn:constraint-unitvariance}, and \refeqn{eqn:apply-ideal-scale}, respectively.
The observation data were generated by mixing clean speech and BG noise. To represent multiple scenarios in different SNRs, three multipliers such as 1, 2, and 4 were applied to the BG noise. We refer to these values as {\it BG multipliers}.
The BF output $y(t)$ was generated as explained in \refsec{sec:optimal-mask};
one or two masks were employed and the BF filter was estimated depending on the BF method, such as \refeqn{eqn:mask-mwf}, \refeqn{eqn:max-snr-gev}, \refeqn{eqn:min-nor-gev}, and \refeqn{eqn:max-sor-gev}.
Given that microphone \#5 is the closest to the speaker, we set $k$ to 5 in \refeqn{eqn:optimal-mask-def} and \refeqn{eqn:ideal-scaling}.  
The backpropagation based on the gradient descent algorithm was utilized only in the first experiment to obtain the optimal mask for each utterance. The estimation of the BF filter and output was implemented in PyTorch \cite{Paszke2019-dm}, which supports the backpropagation of matrix operations in the complex number domain.

\reftable{tab:obs-snr} presents average SNR scores of the observation of microphone \#5 for each BG multiplier.

\begin{table}[b]\begin{center}
\caption{SNR scores {[\rm dB]} of the observation of microphone \#5}
\label{tab:obs-snr}

\begin{tabular}{rrr} \hline
\multicolumn{3}{c}{BG multiplier} \\
\multicolumn{1}{c}{$\times$1} & \multicolumn{1}{c}{$\times$2} & \multicolumn{1}{c}{$\times$4} \\\hline
7.540  & 1.536  & -4.421  \\\hline
\end{tabular}
\end{center}\end{table}

For the evaluation score, we adopted the source-to-distortion ratio (SDR) \cite{Vincent2006-pg}, which is provided as a performance score in this dataset, considering that this score basically represents the closeness of the BF output to the target.

\subsection{Experiment 1: Exploring the peak performance for each BF method}
First, to explore the peak performance of each BF method, we obtained the optimal mask for each utterance, which is the solution to \refeqn{eqn:optimal-mask-def} under the constraints represented as \refeqn{eqn:constraint-nonnegative} and \refeqn{eqn:constraint-unitvariance}, using the backpropagation of 500 iterations.
\reftable{tab:potential-peak} presents SDR scores of the mask-based BFs, including those of the ideal MWF BF using $k=5$ in \refeqn{eqn:ideal-mwf}.
All the mask-based BFs practically demonstrated identical scores comparable to the ideal MWF BF. These are remarkable results, given that they were not theoretically evident, as mentioned in Sections~\ref{subsec:mwf} and \ref{subsec:max-snr}.

\begin{table}[b]\begin{center}
\caption{Peak performance for each method and scenario in SDR {\rm [dB]} compared with that of the ideal MWF.}
\label{tab:potential-peak}

\begin{tabular}{c|c|rrr} \hline
BF        & Mask(s) & \multicolumn{3}{c}{BG multiplier} \\
Method    & used    & \multicolumn{1}{c}{$\times$1} & \multicolumn{1}{c}{$\times$2} & \multicolumn{1}{c}{$\times$4} \\\hline\hline
Max-SNR   & $m_{s},m_{n}$ & 19.430  & {\bf 14.276}  & {\bf 9.451}  \\
Max-SOR   & $m_{s}$          & 19.426  & {\bf 14.276}  & {\bf 9.451}  \\
Min-NOR   & $m_{n}$          & 19.434  & {\bf 14.276}  & {\bf 9.451}  \\
MWF       & $m_{s}$          & 19.438  & 14.268  & 9.430  \\\hline
Ideal MWF &                  & {\bf 19.441} & {\bf 14.276} & {\bf 9.451} \\\hline
\end{tabular}
\end{center}\end{table}

\subsection{Experiment 2: Verifying whether the optimal mask is common for all the BF methods}
Furthermore, to verify whether the optimal mask is common for all the BF methods, we applied the optimal mask obtained in a BF method to another one.
If the same score is achieved after applying the mask, we can claim that the optimal mask is the same between these two BF methods.
Although 12 permutations are possible from four methods, we examined six considered to be nontrivial.
In the cases that $m_{s}(t)$ was applied to a BF method using $m_{n}(t)$, and vice versa, we converted the mask values using the following rules to satisfy \refeqn{eqn:sum-is-constant}:
\begin{eqnarray}
    m_{s}(t) &=& \max_{t}\left\{m_{n}(t)\right\} - m_{n}(t),
        \label{eqn:ms-from-mn}\\
    m_{n}(t) &=& \max_{t}\left\{m_{s}(t)\right\} - m_{s}(t),
        \label{eqn:mn-from-ms}
\end{eqnarray}
where $\max_{t}\left\{\cdot\right\}$ denotes the maximum value over $t$.

The obtained results are presented in \reftable{tab:performance-across-bf-methods}.
The first and second rows indicate that the optimal masks obtained in the max-SNR BF were applied to the max-SOR and min-NOR BFs, respectively. These two rows present lower scores than Experiment 1. This suggests that the optimal mask obtained in the max-SNR BF is not necessarily optimal for the max-SOR or min-NOR BFs.

The third row indicates that the optimal mask $m_{s}(t)$ obtained in the max-SOR BF was applied to the min-NOR BF using \refeqn{eqn:mn-from-ms}, and the fourth row indicates that the mask $m_{n}(t)$ obtained in the min-NOR BF was applied to the max-SOR BF using \refeqn{eqn:ms-from-mn}.
The two rows practically demonstrated the same scores as the max-SOR and min-NOR BFs in \reftable{tab:potential-peak}.
This suggests that the optimal mask can be converted between the max-SOR and min-NOR BFs, despite the fact that the two masks are not the same.

The fifth and sixth rows indicate that the optimal mask $m_{s}(t)$ obtained in a method was applied to the other for the max-SOR and MWF BFs. These two rows demonstrated considerably lower scores than the results of the max-SOR and MWF BFs in \reftable{tab:potential-peak}. These results suggest that the optimal mask is not common between these two BF methods.

\begin{table}[b]\begin{center}
\caption{Extraction performance in SDR {[\rm dB]} after applying the optimal mask to another BF method}
\label{tab:performance-across-bf-methods}
\begin{tabular}{cc|c|rrr} \hline
BF for mask & Applied & Mask & \multicolumn{3}{c}{BG multiplier} \\
estimation & to & used & \multicolumn{1}{c}{$\times$1} & \multicolumn{1}{c}{$\times$2} & \multicolumn{1}{c}{$\times$4} \\\hline\hline
Max-SNR & Max-SOR & $m_{s}$ & 17.084  & 13.493  & 9.165  \\
        & Min-NOR & $m_{n}$ & 18.399  & 14.032  & 9.381  \\\hline
Max-SOR & Min-NOR & $m_{s}$ \& \refeqn{eqn:mn-from-ms} & 19.426  & {\bf 14.276}  & {\bf 9.451}  \\
Min-NOR & Max-SOR & $m_{n}$ \& \refeqn{eqn:ms-from-mn} & {\bf 19.434}  & 14.274  & {\bf 9.451}  \\\hline
Max-SOR & MWF & $m_{s}$ & 15.465  & 10.259  & 5.337  \\
MWF & Max-SOR & $m_{s}$ & 14.230  & 12.092  & 8.458  \\\hline
\end{tabular}
\end{center}\end{table}

\subsection{Experiment 3: Examining whether the IRM can achieve the peak performance}
The third experiment is for examining if the IRM can achieve the peak performance for each BF method.
Considering that the term IRM was ambiguously employed in the related studies \cite{Erdogan2016-jc,Tu2017-ey,Wang2018-yy,Tu2019-qp,Chakrabarty2019-th}, we define several masks that utilize the ratios of the target and interferences based on \cite{Wang2018-yy}.

The IRMs for the target and interferences are defined as
\begin{eqnarray}
m_{s}(t) &=& \left(\frac{\abs{s_{k}(t)}^{2}}{\abs{s_{k}(t)}^{2}+\abs{n_{k}(t)}^{2}}\right)^{\beta}, \label{eqn:def-irm-ms} \\
m_{n}(t) &=& \left(\frac{\abs{n_{k}(t)}^{2}}{\abs{s_{k}(t)}^{2}+\abs{n_{k}(t)}^{2}}\right)^{\beta}, \label{eqn:def-irm-mn}
\end{eqnarray}
where $k=5$ and $\beta=1$ or $0.5$. As another type of ratio mask, we employed the spectral magnitude masks (SMMs) defined as
\begin{eqnarray}
m_{s}(t) &=& \frac{\abs{s_{k}(t)}}{\abs{s_{k}(t) + n_{k}(t)}}, \label{eqn:def-smm-ms} \\
m_{n}(t) &=& \frac{\abs{n_{k}(t)}}{\abs{s_{k}(t) + n_{k}(t)}}. \label{eqn:def-smm-mn}
\end{eqnarray}
In the single-channel TF masking, these masks are ideal in terms of the magnitudes of the target and interferences \cite{Wang2018-yy}.

The obtained results are presented in \reftable{tab:performance-irm-and-smm}. In this table, the combination of the max-SOR and IRM with $\beta=0.5$ demonstrated the best score for all the scenarios. However, these scores were lower than that in Experiment 1. The results suggest that the IRM does not achieve the peak performance for any BF method examined in this study; hence, it differs from the optimal mask for each method.

In addition, \reftable{tab:performance-irm-and-smm} indicates that the max-SNR, max-SOR, and min-NOR BFs present the same scores for the IRM with $\beta=1$. This is because this mask type evidently satisfies both \refeqn{eqn:cov_summation} and \refeqn{eqn:sum-is-constant}. Hence, these three BFs are equivalent in this case, as mentioned in \refsec{subsec:max-snr}.

\begin{table}[b]\begin{center}
\caption{Extraction performance in SDR {[\rm dB]} when the IRM ($\beta=1$, $0.5$) and SMM were employed for each BF method.}
\label{tab:performance-irm-and-smm}
\begin{tabular}{c|l|c|rrr} \hline
BF & Type of & Mask(s) & \multicolumn{3}{c}{BG multiplier} \\
method & ideal mask & used & \multicolumn{1}{c}{$\times$1} & \multicolumn{1}{c}{$\times$2} & \multicolumn{1}{c}{$\times$4} \\\hline\hline
        & IRM ($\beta=1$)   &               & 18.642  & 13.889  & 9.226  \\
Max-SNR & IRM ($\beta=0.5$) & $m_{s},m_{n}$ & 18.313  & 13.790  & 9.203  \\
        & SMM               &               & 17.973  & 13.529  & 9.060  \\\hline
        & IRM ($\beta=1$)   &               & 18.642  & 13.889  & 9.226  \\
Max-SOR & IRM ($\beta=0.5$) & $m_{s}$       & {\bf 18.725}  & {\bf 13.948}  & {\bf 9.275}  \\
        & SMM               &               & 13.640  & 11.447  & 8.147  \\\hline
        & IRM ($\beta=1$)   &               & 18.642  & 13.889  & 9.226  \\
Min-NOR & IRM ($\beta=0.5$) & $m_{n}$       & 18.267  & 13.747  & 9.160  \\
        & SMM               &               & 17.372  & 12.735  & 8.103  \\\hline
        & IRM ($\beta=1$)   &               & 17.316  & 12.788  & 8.375  \\
MWF     & IRM ($\beta=0.5$) & $m_{s}$       & 16.228  & 11.799  & 7.477  \\
        & SMM               &               & 17.109  & 12.316  & 7.734  \\\hline
\end{tabular}
\end{center}\end{table}
\section{Discussion} \label{sec:discussion}

In this section, we discuss the experimental results for each issue mentioned in \refsec{sec:framework}.

\subsection{Discussion on Issue 1}
\label{subsec:discuss-issue1}

The results of Experiment 1 provide the answer to Issue 1. The peak performance of each BF method is practically identical and comparable to the upper bound given by the ideal MWF BF. These findings differ from the conventional idea presented in \reffig{fig:conventional_image} and may deprive the meaning of the discussion on the BF that can achieve the best extraction performance. However, it is an open question whether these findings are applicable to any BF method and dataset. Therefore, further investigation is required.

\subsection{Discussion on Issue 2}
\label{subsec:discuss-issue2}

The results of Experiment 2 provide the answer to Issue 2. Although the optimal mask is not common, it depends on the BF method adopted.
This finding differs from the conventional concept presented in \reffig{fig:conventional_image}, similar to the discussion on Issue 1.
However, we consider this to be natural because the optimal mask is formulated as the solution to a different problem for each BF method, as explained in \refsec{sec:optimal-mask}.

In addition, \reftable{tab:performance-across-bf-methods} indicates several remarkable points. Although both the max-SOR and min-NOR BFs are derived from the max-SNR BF, as mentioned in \refsec{subsec:max-snr}, the top two rows suggest that $m_{s}(t)$ and $m_{n}(t)$ are optimal for the max-SNR BF, while not for the max-SOR or min-NOR BFs. We consider that the reason for this is that the optimal masks obtained in the max-SNR BF do not satisfy \refeqn{eqn:cov_summation} or \refeqn{eqn:sum-is-constant}. Hence, the max-SNR BF is not equivalent to the other two BFs in this case. In contrast, the third and fourth rows in this table suggest that the max-SOR and min-NOR BFs are equivalent if both $m_{s}(t)$ and $m_{n}(t)$ satisfy \refeqn{eqn:sum-is-constant}. Moreover, the bottom two rows exhibited lower scores than the other rows. Although $m_{s}(t)$ has commonly been interpreted as the mask for the target, the facts suggest that the optimal mask $m_{s}(t)$ for the max-SOR BF differs significantly from that of the MWF BF and vice versa. 

\begin{figure}[t]
  \centering
  \includegraphics[width=0.6\linewidth]{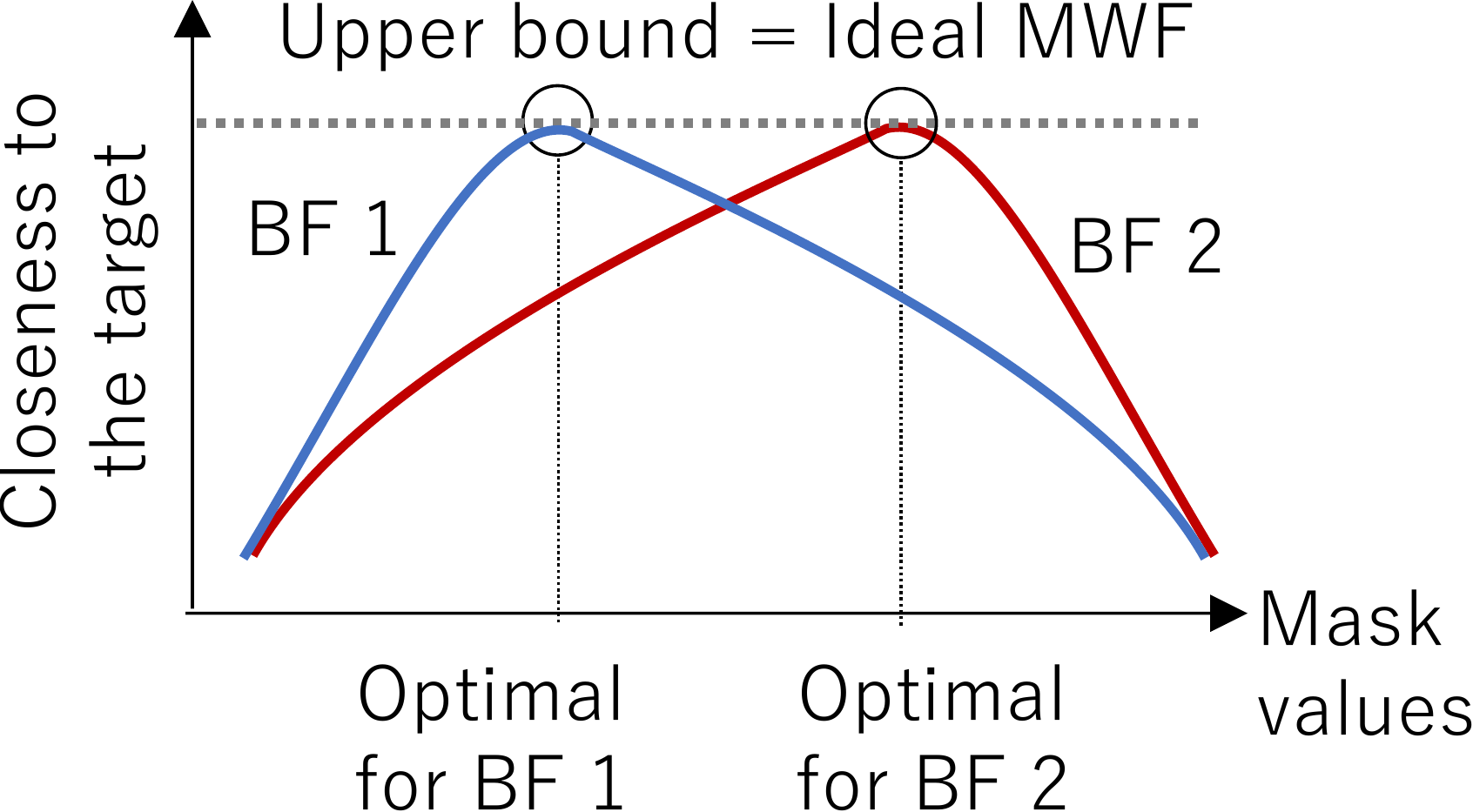}
  \caption{Obtained concept of the peak performance and optimal mask among multiple BF methods}
  \label{fig:discovered_image}
\end{figure}

From the discussion on Issues 1 and 2, we have obtained a novel idea for the peak performance and optimal mask among multiple BF methods, as illustrated in \reffig{fig:discovered_image}. This indicates that the optimal mask depends on the BF method adopted, while the peak performance is the same among multiple BF methods and comparable to the upper bound achieved by the ideal MWF.

\subsection{Discussion on Issue 3}
\label{subsec:discuss-issue3}
The results of Experiment 3 provide the answer to Issue 3. 
Although these masks have been considered to be optimal for any BF method, \reftable{tab:performance-irm-and-smm} suggests that the IRM and SMM do not achieve the upper-bound performance and differ from the optimal mask for any BF method.
These findings can explain why the studies presented in \reftable{tab:comparison} mentioned a different BF method as the best.
\reftable{tab:performance-irm-and-smm} indicates that the BF method that appears the best depends on the mask type employed. For example, focusing on the IRM with $\beta=0.5$ in this table, we can determine that the max-SOR BF performs the best. However, this result only suggests that this mask type is the closest to the optimal mask for the max-SOR in this dataset.

Furthermore, the fact that the IRM is not optimal imposes another issue on us. This is how the optimal mask is interpreted and represented as a formula. However, this is also an open question.

\section{Conclusions} \label{sec:conclusions}

In this study, we investigated mask-based BFs such as the max-SNR, max-SOR, min-NOR, and MWF BFs.
To explore the peak performance for each BF method, we obtained the optimal mask for each utterance by minimizing the MSE between the BF output and target.
We experimentally verified that these four methods have the same peak performance as the upper bound provided by the ideal MWF BF.
Via additional experiments that applied the optimal mask across BF methods, we determined that the optimal mask differed for the BF method used. However, the mask values can be converted between the max-SOR and min-NOR BFs.
These findings differed from the conventional idea that the optimal mask would be common and the peak performance would depend on the BF method.
We verified that the IRM did not achieve the peak performance for these four BFs. Hence, this mask type was not optimal.
We expect that these findings will contribute to the improvement of mask-based BFs. 

Given that these findings are currently experimental, in the future, we would attempt to establish their theoretical background and investigate whether these apply to other BF methods and datasets.

The experimental system has been shared in \url{https://github.com/hreshare/optimal_beamformers/}.

\printbibliography

\end{document}